\documentclass{article}
\usepackage{tikz,bm,amsmath,caption,subcaption}
\usetikzlibrary{arrows,patterns}
\usepackage{amsthm}
\usepackage{fullpage,varwidth}
\usepackage{amssymb}
\usepackage{url}
\usepackage{cite}
\tikzset{transform shape}
\usetikzlibrary{knots}
\usepackage{graphicx}
\usepackage[backref=page]{hyperref}

\usepackage{hyperref}
\hypersetup{
    colorlinks=true, 
    linktoc=all,     
    linkcolor=blue,  
}

\def\eqref#1{{\rm(\ref{#1})}}

\newcommand{\cP}{\mathcal{P}}

\newcommand{\R}{\mathbb{R}}
\newcommand{\Z}{\mathbb{Z}}
\newcommand{\N}{\mathbb{N}}
\newcommand{\Q}{\mathbb{Q}}
\newcommand{\C}{\mathbb{C}}

\newcommand{\Var}{\mathrm{Var}}

\newcommand{\NP}{\mathcal{NP}}

\newcommand{\F}{\mathbb{F}}
\newcommand{\VP}{\mathcal{VP}}
\newcommand{\VNP}{\mathcal{VNP}}

\newcommand{\poly}{\mathrm{poly}}

\newtheoremstyle{cited}%
  {3pt}
  {3pt}
  {\itshape}
  {}
  {\bfseries}
  {}
  {.5em}
  {\thmname{#1} \thmnumber{#2} \thmnote{\normalfont#3}}
\newtheoremstyle{defcited}%
  {3pt}
  {3pt}
  {\upshape}
  {}
  {\bfseries}
  {.}
  {.5em}
  {\thmname{#1} \thmnumber{#2} \thmnote{\normalfont#3}}

\newtheorem{theorem}{Theorem}[section]
\newtheorem{corollary}[theorem]{Corollary}

\theoremstyle{definition}
\newtheorem{definition}[theorem]{Definition}

\newtheorem{conjecture}[theorem]{Conjecture}

\theoremstyle{cited}
\newtheorem{citedthm}[theorem]{Theorem}

\theoremstyle{defcited}

\title{Interactions of Computational Complexity Theory and Mathematics}
\author{Avi Wigderson}

\begin{document}
\maketitle

\begin{abstract}

[This paper is a (self contained) chapter in a new book on computational complexity theory, called {\em Mathematics and Computation}, whose draft is available at  \url{https://www.math.ias.edu/avi/book}].
\vspace{.1 in}
\noindent

We survey some concrete interaction areas between computational complexity theory and different fields of mathematics. We hope to demonstrate here that hardly any area of modern mathematics is untouched by the computational connection (which in some cases is completely natural and in others may seem quite surprising). In my view, the breadth, depth, beauty and novelty of these connections is inspiring, and speaks to a great potential of future interactions (which indeed, are quickly expanding). We aim for {\em variety}. We give short, simple descriptions (without proofs or much technical detail) of ideas, motivations, results and connections; this will hopefully entice the reader to dig deeper. Each vignette focuses only on a single topic within a large mathematical filed. We cover the following:

\begin{itemize}
\setlength\itemsep{.2 em}
\item Number Theory: {\em Primality testing}
\item Combinatorial Geometry: {\em Point-line incidences}
\item Operator Theory: {\em The Kadison-Singer problem}
\item Metric Geometry: {\em Distortion of embeddings}
\item Group Theory: {\em Generation and random generation}
\item Statistical Physics: {\em Monte-Carlo Markov chains}
\item Analysis and Probability: {\em Noise stability}
\item Lattice Theory: {\em Short vectors}
\item Invariant Theory: {\em Actions on matrix tuples}
\end{itemize} 

\end{abstract}

\newpage
\section{introduction}

The Theory of Computation (ToC) lays out the mathematical foundations of computer science.  
I am often asked if ToC is a branch of Mathematics, or of Computer Science. The answer is easy: it is clearly both (and in fact, much more). Ever since
Turing's 1936 definition of the {\em Turing machine}, we have had a formal mathematical model of computation that enables the rigorous mathematical study of computational tasks, algorithms to solve them, and the resources these require. At the same time, the simple description of the Turing machine allowed its simple logical structure to be implemented in hardware, and its universal applicability fueled the rapid development of computer technology, which now dominates our life. 
 
Computation was part mathematics from its origins, and motivated many of its developments. Algorithmic questions have occupied mathematicians throughout history (as elaborated in the introduction to the book~\cite{Wig17}), and this naturally grew considerably when computers arrived. However, the advent of {\em computational complexity theory} over the past few decades has greatly expanded and deepened these connections. The study of new diverse models generated and studied in complexity theory broadened the nature of mathematical problems it encountered and formulated, and the mathematical areas and tools which bear upon these problems. This expansion has led to numerous new interactions that enrich both disciplines. This survey tells the stories of some of these interactions with different mathematical fields, illustrating their diversity. 

We note in passing that a similar explosion of connections and interactions is underway between ToC and practically {\em all} sciences. These stem from computational aspects of diverse natural processes, which beg for algorithmic modeling and analysis. As with mathematics, these interactions of ToC with the sciences enrich both sides,
expose {\em computation} as a central notion of intellectual thought, and highlights its study as an independent discipline, whose mission and goals expand way beyond those emanating from its parent fields of Math and CS. But this is the subject of a different survey (which I partly provide in the last chapter of~\cite{Wig17}).

Back to the interactions of computational complexity theory and different areas of math.
I have chosen to focus on essentially one problem or development within each mathematical field. Typically this touches only a small subarea, and does not do justice to a wealth of other connections. Thus each vignette should be viewed as a demonstration of a larger body of work and even bigger potential. Indeed, while in some areas the collaborations are quite well established, in others they are just budding, with lots of exciting problems waiting to be solved and theories to be developed. Furthermore, the connections to algorithms and complexity (which I explain in each) are quite natural in some areas, but quite surprising in others. While the descriptions of each topic are relatively short, they include background and intuition, as well as further reading material. Indeed, I hope these vignettes will tempt the reader to explore further.

Here is a list of the covered areas and topics chosen in each; these sections can be read in any order. The selection of fields and foci is affected by my personal taste and limited knowledge. More connections to other fields like Combinatorics, Optimization, Logic, Topology and Information Theory appear in parts of the book~\cite{Wig17}. 

\begin{itemize}
\item Number Theory: {\em Primality testing}
\item Combinatorial Geometry: {\em Point-line incidences}
\item Operator Theory: {\em The Kadison-Singer problem}
\item Metric Geometry: {\em Distortion of embeddings}
\item Group Theory: {\em Generation and random generation}
\item Statistical Physics: {\em Monte-Carlo Markov chains}
\item Analysis and Probability: {\em Noise stability}
\item Lattice Theory: {\em Short vectors}
\item Invariant Theory: {\em Actions on matrix tuples}
\end{itemize}

\section{Number Theory}\label{number-theory}

As mentioned, the need to efficiently compute mathematical objects has been central to mathematicians and scientists throughout history, and of course the earliest subject is arithmetic.
Perhaps the most radical demonstration is the place value system we use to represent integers, which is in place for Millenia precisely due to the fact that it supports extremely efficient manipulation of arithmetic operations. The next computational challenge in arithmetic, since antiquity, was accessing the multiplicative structure of integers represented this way.

Here is an except from C.\ F.\ Gauss' appeal\footnote{Which is of course in Latin. I copied this English translation from a wonderful survey of Granville~\cite{Gra05} on the subject matter of this section.} to the mathematics community of his time (in article 329 of \emph{Disquisitiones Arithmeticae} (1801)), regarding the computational complexity of {\em testing primality} and {\em integer factorization}. The importance Gauss assigns to this computational challenge, his frustration of the state of art, and his imploring the mathematical community to resolve it shine through!

{\it The problem of distinguishing prime numbers from composite
numbers, and of resolving the latter into their prime factors is known to be
one of the most important and useful in arithmetic. It has engaged the industry
and wisdom of ancient and modern geometers to such an extent that it would
be superfluous to discuss the problem at length. Nevertheless we must confess
that all methods that have been proposed thus far are either restricted to very
special cases or are so laborious and difficult that even for numbers that do
not exceed the limits of tables constructed by estimable men, they try the
patience of even the practiced calculator. And these methods do not apply at
all to larger numbers \dots{}  the dignity of the science itself
seems to require that every possible means be explored for the solution of a
problem so elegant and so celebrated.}

We briefly recount the state-of-art of these two basic algorithmic problems in number theory.
A remarkable response to Gauss' first question, {\em efficiently deciding primality}, was found in 2002 by Agrawal, Kayal, and Saxena~\cite{AKS04}. The use of symbolic polynomials for this problem is completely novel. Here is their elegant characterization of prime numbers.

\begin{citedthm}[\cite{AKS04}]\label{Primality}

An integer  $N\geq 2$ is prime if and only if
\begin{itemize}
\item $N$ is not a perfect power,
\item $N$ does not have any prime factor $\leq (\log N)^4$,
\item For every $r,a < (\log N)^4$ we have the following equivalence of polynomials over $\Z_N[X]$:
$$(X+a)^N \equiv X^N +a \mod (X^r-1)$$
\end{itemize}
\end{citedthm}

It is not hard to see that this characterization gives rise to a simple algorithm for testing primality that is deterministic, and runs in time that is {\em polynomial} in the binary description length of $N$. Previous deterministic algorithms either assumed the generalize Riemann hypothesis \cite{Mil76} or required slightly superpolynomial time \cite{APR83}.
The AKS deterministic algorithm came after a sequence of efficient {\em probabilistic} algorithms \cite{SoSt77, Rab80, GoKi86, AdHu92}, 
some elementary and some requiring sophisticated use and development of number theoretic techniques. These probabilistic and deterministic algorithms were partly motivated by, and are important to the field of cryptography. 

What is not so well-known, even for those who did read the beautiful, ingenious proof in~\cite{AKS04},  is that AKS developed their deterministic algorithm by carefully ``de-randomizing'' a previous probabilistic algorithm for primality of \cite{AgBi03} (which uses polynomials). 
We note that {\em de-randomization}, the conversion of probabilistic algorithms into deterministic ones, is by now a major area in computational complexity with a rich theory, and many other similar successes as well as challenges. The stunning possibility that {\em every} efficient probabilistic algorithm has a deterministic counterpart is one of the major problems of computational complexity, and there is strong evidence supporting it (see~\cite{ImWi97}). Much more on this can be found in the randomness chapters of~\cite{Wig17}.

Gauss' second challenge, of whether efficiently factoring integers is possible, remains open. But this very challenge has enriched computer science, both practical and theoretical in several major ways. Indeed, the assumed hardness of factoring is the main guarantee of security in almost all cryptographic and e-commerce systems around the world (showing that difficult problems can be useful!). More generally, cryptography is an avid consumer of number theoretic notions, including elliptic curves, Weil pairings, and more, which are critical to a variety of cryptographic primitives and applications. These developments shatter Hardy's view of number theory as a completely useless intellectual endeavor.

There are several problems on integers whose natural definitions depend on factorization, but can nevertheless be solved efficiently, bypassing the seeming need to factor. Perhaps the earliest algorithm ever formally described is Euclid's algorithm for computing the GCD (greatest common divisor) of two given integers\footnote{It extends to polynomials, and allows efficient way of computing multiplicative inverses in quotient rings of $\Z$ and $\F[x]$.} $m$ and $n$. Another famous such algorithm is for computing the Legendre-Jacobi symbol $(\frac{m}{n})$ via Gauss' law of quadratic reciprocity.

A fast algorithm for factoring may come out of left-field with the new development of quantum computing, the study of computers based on quantum-mechanical principles, which we discussed in the quantum chapter of the book~\cite{Wig17}. Shor has shown in \cite{Sho94} that such computers are capable of factoring integers in polynomial time. This result led governments, companies, and academia to invest billions in developing technologies which will enable building large-scale quantum computers, and the jury is still out on the feasibility of this project. There is no known theoretical impediment for doing so, but one possible reason for failure of this project is the existence of yet-undiscovered principles of quantum mechanics.

Other central computational problems include solving polynomial equations in finite fields, for which one of the earliest efficient (probabilistic) algorithm was developed by Berlekamp~\cite{Ber67} (it remains a great challenge to de-randomize this algorithm!). Many other examples can be found in the Algorithmic Number Theory book \cite{BaSh97}.

\section{Combinatorial geometry}

What is the smallest area of a planar region which contains a unit length segment in {\em every} direction? This is the Kakeya needle problem (and such sets are called {\em Kakeya sets}), which was solved surprisingly by Besicovich \cite{Bes19}
who showed that this area can be arbitrarily close to zero! Slight variation on his method produces a 
Kakeya set of Lebesque measure zero. It makes sense to replace ``area'' (namely, Lesbegue measure) by the more robust measures, such as the Hausdorff and Minkowski dimensions. This changes the picture: Davies~\cite{Dav71} proved that a Kakeya set in the plane must have full dimension (=2) in both measures, despite being so sparse in Lebesgue measure.

It is natural to extend this problem to higher dimensions. However, obtaining analogous results (namely, that the Hausdorff and Minkowski dimensions are full) turns out to be extremely difficult.
Despite the seemingly recreational flavor, this problem has significant importance in a number of mathematical areas (Fourier analysis, Wave equations, analytic number theory, and randomness extraction), and has been attacked through a considerable diversity of mathematical ideas (see \cite{Tao09}).

The following finite field analogue of the above Euclidean problem was suggested by Wolff \cite{Wol99}.
Let $\F$ denote a finite field of size $q$. A set $K\subseteq \F^n$ is called Kakeya if it contains a line in every direction. More precisely, for every direction $b\in \F^n$ there is a point $a\in \F^n$ such that the line $\{ a+bt \, :\, t\in \F\}$ is contained in $K$. As above, we would like to show that any such $K$ must be large (think of the dimension  $n$ as a large constant, and the field size $q$ as going to infinity).

\begin{conjecture}\label{Wolff}
Let $K \subseteq \F^n$ be a
Kakeya set. Then $|K| \geq C_n  q^n$,
where $C_n$ is a constant depending only on the dimension $n$.
\end{conjecture}

The best exponent of $q$ in such a lower bound intuitively corresponds to the Hausdorff and Minkowski dimensions in the Euclidean setting. Using sophisticated techniques from arithmetic combinatorics, Bourgain, Tao and others improved the trivial bound of $n/2$ to about $4n/7$. 

Curiously, the exact same conjecture arose, completely independently, within ToC, from the work~\cite{LRVW03} on {\em randomness extractors},  
an area which studies the ``purification'' of ``weak random sources''
(see e.g.\ the survey \cite{Vad11} on this important notion). In~\cite{LRVW03} Wolff's conjecture takes a probabilistic form, asking about the (min)-entropy of a random point on a random line in a Kakeya set.
With this motivation, Dvir~\cite{Dvi09} brilliantly proved the Wolff conjecture (sometimes called the Finite Field Kakeya conjecture), using the (algebraic-geometric) ``polynomial method'' (which is inspired by techniques in decoding algebraic error-correcting codes). Many other applications of this technique to other geometric problems quickly followed, including the Guth-Katz~\cite{GuKa10}
resolution of the famous Erd\H{o}s distance problem, as well as for optimal randomness extraction and more (some are listed in Dvir's survey~\cite{Dvi10}). 

Subsequent work determined the exact value of the constant $C_n$ above (up to a factor of 2)~\cite{DKSS13}.

\begin{citedthm}[\cite{DKSS13}]\label{DKSS13}
Let $K \subseteq \F^n$ be a
Kakeya set. Then $|K| \geq   (q/2)^n$. On the other hand, there exist Kakeya sets of size $\leq 2\cdot (q/2)^n$.
\end{citedthm}

Many other problems regarding incidences of points and lines (and higher-dimensional geometric objects) have been the source of much activity and collaboration between geometers, algebraists, combinatorialists and computer scientists. The motivation for these questions in the computer science side come from various sources, e.g.\ problems on local correction of errors \cite{BDWY13} 
and de-randomization \cite{DvSh07,KaSa09}. 
Other incidence theorems, e.g.\ Szemer{\'e}di-Trotter~\cite{SzTr83}  
and its finite field version of Bourgain-Katz-Tao~\cite{BKT04}
have been used e.g.\ in randomness extraction \cite{BIW06} 
and compressed sensing~\cite{GLR10}.

\section{Operator theory}\label{operator}

The following basic mathematical problem of Kadison and Singer from 1959 \cite{KaSi59} was intended to formalize a basic question of Dirac concerning the ``universality'' of measurements in quantum mechanics. We need a few definitions. Consider $B({\cal H})$, the algebra of continuous linear operators on a Hilbert space ${\cal H}$. Define a {\em state} to be a linear functional $f$ on $B({\cal H})$, normalized to $f(I)=1$, which takes non-negative values on positive semidefinite operators. The states form a convex set, and a state is called {\em pure} if it is not a convex combination of other states. Finally, let $D$ be the sub-algebra of  $B({\cal H})$ consisting of all {\em diagonal} operators (after fixing some basis). 

Kadison and Singer asked if every pure state on $D$ has a {\em unique} extension to $B({\cal H})$. This problem on infinite-dimensional operators found a host of equivalent formulations in finite dimensions, with motivations and intuitions from operator theory, discrepancy theory,  Banach space theory, signal processing, and probability. All of them were solved affirmatively in recent work of Marcus, Spielman, and Srivastava \cite{MSS13a} (which also surveys the many related conjectures). Here is one statement they prove, which implies the others.

\begin{citedthm}[\cite{MSS13a}]\label{MSS13a}
For every $\epsilon>0$, there is an integer $k=k(\epsilon)$ so that the following holds. Fix any $n$ and any $n\times n$ matrix $A$ with zeros on the diagonal and of spectral norm 1. Then there is a partition of $\{1, 2, \cdots, n\}$ into $k$ subsets, $S_1, S_2, \cdots, S_k$, so that each of the principal minors $A_i$ (namely $A$ restricted to rows and columns in $S_i$) has spectral norm at most $\epsilon$.
\end{citedthm}

This statement clearly implies that one of the minors has linear size, at least $n/k$. This consequence is known as the {\em Restricted Invertibility} Theorem of Bourgain and Tzafriri \cite{BoTz91}, itself an important result in operator theory. 

How did computer scientists get interested in this problem? Without getting into too many details, here is a sketchy description of the meandering path which led to this spectacular result. 

A central computational problem, at the heart of numerous applications, is solving a linear system of equations. While Gaussian elimination does the job quite efficiently (the number of arithmetic operations is about $n^3$ for $n\times n$ matrices), for large $n$ this is still inefficient. Thus faster methods are sought, hopefully nearly linear in the number of non-zero entries of the given matrix. For {\em Laplacian}\footnote{Simply, symmetric PSD matrices with zero row sum.} linear systems (arising in many graph theory applications, such as computing electrical flows and random walks), Spielman and Teng \cite{SpTe11} achieved precisely that! A major notion they introduced was {\em spectral sparsifiers} of matrices (or equivalently, weighted graphs). 

A sparsifier of a given matrix is another matrix, with far fewer (indeed, linear) non-zero entries, which nevertheless has essentially the same (normalized) spectrum as the original (it is not even obvious that such a sparse matrix exists). We note that a very special case of sparsifiers of complete graphs are by definition {\em expander graphs}\footnote{All non-trivial eigenvalues of the complete graph (or constant matrix) are 0, and an expander is a sparse graph in which all non-trivial eigenvalues are tiny.} (see much more about this central concept of expanders  in~\cite{HLW06,Wig17}).  The algorithmic applications led to a quest for optimal constructions of sparsifiers for arbitrary Laplacian matrices (in terms of trade-off between sparsity and approximation), and these were beautifully achieved in~\cite{BSS14} (who also provided a deterministic polynomial time algorithm to construct such sparsifiers). This in turn has led~\cite{SpSr12} to a new proof, with better analysis, of the Restricted Invertibility theorem mentioned above, making the connection to the Kadison-Singer problem. 

However, the solution to Kadison-Singer seemed to require another detour. The same team~\cite{MSS13} first resolved a bold conjecture of Bilu and Linial~\cite{BiLi06} on the spectrum of ``signings'' of matrices\footnote{Simply, this beautiful conjecture states that for {\em every} $d$-regular graph, there exist $\{-1,1\}$ signs of the edges which make all eigenvalues of the resulting signed adjacency matrix lie in the ``Ramanujan interval'' $[-2\sqrt{d-1},2\sqrt{d-1}]$.}. This conjecture was part of a plan for a {\em simple}, iterative construction of Ramanujan graphs, the best\footnote{With respect to the spectral gap. This is one of a few important expansion parameters to optimize.} possible  expander graphs. Ramanujan graphs were introduced and constructed in \cite{LPS88, Mar88}, but rely on deep results in number theory and algebraic geometry (believed by some to be essential for {\em any} such construction). Bilu and Linial sought instead an elementary construction, and made progress on their conjecture, showing how their iterative approach gives yet another way to construct ``close to'' Ramanujan expanders.

To prove the Bilu-Linial conjecture (and indeed produce bipartite Ramanujan graphs of every possible degree---something the algebraic constructions couldn't provide), \cite{MSS13} developed a theory of {\em interlacing polynomials} that turned out to be the key technical tool for resolving Kadison-Singer in~\cite{MSS13a}.  In both cases, the novel view is to think of these conjectures probabilistically, and analyze the norm of a random operator by analyzing the average characteristic polynomial. That this method makes sense and actually works is deep and mysterious. Moreover, it provides a new kind of existence proofs for which no efficient algorithm (even probabilistic) of finding the desired objects is known. The analysis makes heavy use of the theory of {\em Real stable} polynomials, and the inductive process underlying it is reminiscent (and inspired by) Gurvits'~\cite{Gur08} remarkable proof of the van der Waerden conjecture and its generalizations\footnote{This is yet another example of structural result (on doubly stochastic matrices) whose proof was partly motivated by algorithmic ideas. The connection is  the use of hyperbolic polynomials in optimization (more specifically, as barrier functions in interior point methods.}. 

\section{Metric Geometry}

How close one metric space is to another is captured by the notion of {\em distortion}, measuring how distorted distances of one become  when embedded into the other. More precisely,
\begin{definition}
Let $(X,d)$ and $(X',d')$ be two metric spaces. An embedding $f:X \rightarrow X'$ has distortion $\leq c$ if for every pair of points $x,y \in X$ we have $$d(x,y) \leq d'(f(x),f(y)) \leq c\cdot d(x,y).$$ When $X$ is finite and of size $n$, we allow $c=c(n)$ to depend on $n$.
\end{definition}

Understanding the best embeddings between various metric and normed spaces has been a long endeavor in Banach space theory and metric geometry. An example of one major result in this area is Bourgain's embedding theorem~\cite{Bou85}.

\begin{citedthm}[\cite{Bou85}]\label{L2}
Every metric space of size $n$ can be embedded into Euclidean space $L_2$ with distortion $O( \log n)$.
\end{citedthm}

The first connection between these structural questions and computational complexity was made in the important paper of Linial, London and Rabinovich~\cite{LLR95}. They asked for efficient algorithms for actually finding embeddings of low distortion, and noticed that for some such problems it is natural to use semi-definite programming. They applied this geometric connection to get old and new results for algorithmic problems on graphs (in particular, the sparsest cut problem we will soon discuss. Another motivation they discuss (which quickly developed into a major direction in approximation algorithms) is that some computations (e.g. finding nearest neighbors) are more efficient in some spaces than others, and so {\em efficient}, low-distortion embedding may provide useful reductions from harder to easier space. They describe such an efficient algorithm implementing Bourgain's Theorem~\ref{L2} above, and also prove that his bound is best possible (the metric proving it is simply the distances between points in any constant-degree {\em expander} graph\footnote{The presence of such graphs in different sections illustrate how fundamental they are in diverse mathematical areas, and the same holds for algorithms and complexity theory.}).

The next shift in the evolution of this field, and in the level of interactions between geometers and ToC researchers, came from trying to prove ``hardness of approximation'' results. One example is the Goemans-Linial conjecture~\cite{Goe97,Lin02}, studying the sparsest cut problem, about the relation between $L_1$ and the ``negative type'' metric space $L_2^2$ (a general class of metrics which arise naturally in several contexts). Roughly, thes are  metrics on $\R^n$ in which Euclidean distances are squared. More precisely, a metric $(X,d)$  is of negative type (namely, in $L_2^2$),  if $(X,\sqrt{d})$,  is isometric (has no distortion) to a subset of $L_2$.

\begin{conjecture}
Every $L_2^2$ metric can be embedded into $L_1$ with constant distortion.
\end{conjecture}

This conjecture was proved false by Khot and Vishnoi~\cite{KhVi05}:
\begin{citedthm}[\cite{KhVi05}]
For every $n$ there are $n$-point subsets of $L_2^2$ for which every embedding to $L_1$ requires distortion $\Omega(\log\log n)^{1/6}$.
\end{citedthm}

Far more interesting than the result itself is its origin. Khot and Vishnoi were trying to prove that the (weighted) ``sparsest cut'' problem  is hard to approximate. They managed to do so under a computational assumption, known as the {\em Unique Games} conjecture of Khot~\cite{Kho02} via a so-called {\em PCP}-reduction (see also~\cite{Kho10,Wig17}). The elimination of this computational assumption is the magical part, that demonstrates the power and versatility of reductions between computational problems. They apply their PCP reduction  to a {\em particular}, carefully chosen unique games instance, which cannot be well approximated by a certain semi-definite program. The outcome was an instance of the sparsest cut problem which the same reduction ensures is hard to approximate by a semi-definite program. As discussed above, that outcome instance could be understood as a metric space, and the hardness of approximation translates to the required distortion bound! 

The exact distortion of embedding $L_2^2$ into $L_1$ has been determined precisely to be $\sqrt{\log n}$ (up to lower order factors) in two beautiful sequences of works developing new algorithmic and geometric tools; we mention only the final word for each, as these papers contain a detailed history. On the upper bound side, the efficient algorithm approximating non-uniform sparsest cut to a factor $\sqrt{\log n}\log\log n$, which yields the same distortion bound, was obtained by Arora, Lee and Naor~\cite{ALN08} via a combination of the so-called ``chaining argument'' of~\cite{ARV04} and the ``measured descent'' embedding method of~\cite{KLMN05}.
A lower bound of $\sqrt{\log n}$ on the distortion was 
very recently proved by Naor and Young~\cite{NaYo17} using a new isoperimetric inequality on the Heisenberg group.

Another powerful connection between such questions and ToC is through (again) expander graphs. A basic example is that the graph metric of any constant-degree expander proves that Bourgain's embedding theorem above is optimal! Much more sophisticated examples arise from trying to understand (and perhaps disprove) the Novikov and the Baum-Connes conjectures (see \cite{KaYu06}). This program relies on another, much weaker notion of {\em coarse} embedding. 

\begin{definition}
$(X,d)$ has a coarse embedding into $(X',d')$ if there is a map $f:X \rightarrow X'$ and two increasing, unbounded real functions $\alpha, \beta$ such that for every two points $x,y \in X$, $$\alpha(d(x,y)) \leq d'(f(x),f(y)) \leq \beta(d(x,y)).$$
\end{definition}

Gromov~\cite{Gro87} was the first to construct a metric (the word metric of a group) which cannot be coarsely embedded into a Hilbert space. His construction uses an infinite family of {\em Cayley} expanders (graphs defined by groups). This result was greatly generalized  by Lafforgue~\cite{Laf08} and Mendel-Naor \cite{MeNa14}, who constructed graph metrics that cannot be coarsely embedded into any {\em uniformly convex} space. It is interesting that while Lafforgue's method is algebraic, the Mendel-Naor construction follows the combinatorial {\em zig-zag} construction of expanders~\cite{RVW02} from computational complexity.
 
Many other interaction projects regarding metric embeddings and distortion we did not touch on include their use in numerous algorithmic and data structure problems like clustering, distance oracles the $k$-server problem, as well as the fundamental interplay between distortion and  {\em dimension reduction} relevant to both geometry and CS, where so many basic problems are open.

\section{Group Theory}

Group theorists, much like number theorists, have been intrinsically interested in computational problems since the origin of the field. For example, the {\em word problem} (given a word in the generators of some group, does it evaluate to the trivial element?)\ is so fundamental to understanding any group one studies, that as soon as language was created to formally discuss the computational complexity of this problem, hosts of results followed trying to pinpoint that complexity. These include decidability and undecidability results once Turing set up the theory of computation and provided the first undecidable problems, and these were followed with $\NP$-completeness results and efficient algorithms once $\cP$ and $\NP$ were introduced around 1970. Needless to say, these {\em algorithmic results} inform of  {\em structural} complexity of the groups at hand. And the word problem is but the first example. Another demonstration is the beautiful interplay between algorithmic and structural advances over decades, on the {\em graph isomorphism problem}, recently leading to breakthrough of Babai~\cite{Bab15}!  A huge body of work is devoted to finding efficient algorithms for computing commutator subgroups, Sylow subgroups, centralizers, bases, representations, characters, and a host of other important substructures of a group from some natural description of it. Excellent textbooks include~\cite{HEO05,Ser03}. 

Here we focus on two related problems, the {\em generation} and {\em random generation} problems, and new conceptual notions borrowed from computational complexity which are essential for studying them. Before defining them formally (below), let us consider an example. Assume I hand you 10 invertible matrices, say 100 $\times$ 100 in size, over the field of size 3. Can you tell me if they generate another such given matrix? Can you even produce convincing evidence of this before we both perish? How about generating a random matrix in the subgroup spanned by these generators? The problem, of course, is that this subgroup will have size far larger than the number of atoms in the known universe, so its elements cannot be listed, and typical words generating elements in the group may need to be prohibitively long. Indeed, even the extremely special cases, for elements in $\Z_p^*$ (namely one, $1\times 1 $ matrix), the first question is related to the {\em discrete logarithm} problem, and for $\Z_{p\cdot q}^*$ it is related to the {\em integer factoring} problem, both currently requiring exponential time to solve (as a function of the description length).

Let us consider any finite group $G$ and let $n \approx \log |G|$ be roughly the length of a description of an element of $G$. Assume we are given $k$ elements in $G$, $S=\{s_1, s_2, \ldots , s_k\}$. 
It would be ideal if the procedures we describe would work in time polynomial in $n$ and $k$ (which prohibits enumerating the elements of $G$, whose size is exponential in $n$).

The {\em generation problem} asks if  a given element $g\in G$ is generated by $S$. How does one prove such a fact? A standard certificate for a positive answer is a {\em word} in the elements of $S$ (and their inverses) which evaluates to $g$. However, even if $G$ is cyclic, the shortest such word may be exponential in $n$. An alternative, computationally motivated description, is to give a {\em program} for $g$. Its definition shows that the term ``program'' suits it perfectly, as it has the same structure as usual computer programs, only that instead of applying some standard Boolean or arithmetic operations, we use the group operations of multiplication and inverse.

\begin{definition}
A {\em program} (over $S$) is  a finite sequence of elements $g_1, g_2, \cdots , g_m$, where every element $g_i$ is either in $S$, or is the inverse of a previous $g_j$, or is the product of previous $g_j, g_\ell$. We say that it computes $g$ simply if $g=g_m$.
\end{definition} 

In the cyclic case, programs afford exponential savings over words in description length, as a program allows us to write large powers by repeatedly squaring elements. What is remarkable is that such savings are possible for {\em every} group. This discovery of Babai and Szemer{\'e}di~\cite{BaSz84} says that every element of every group has an extremely succinct description in terms of any set of elements generating it.

\begin{citedthm}[\cite{BaSz84}]
For every group $G$, if a subset of elements $S$ generates another element $g$, then there is a program of length at most $n^2 \approx (\log |G|)^2$ which computes $g$ from S.
\end{citedthm}

It is interesting to note that the proof uses a structure which is very combinatorial and counterintuitive for group theorists: that of  a {\em cube}, which we will see again later. For a sequence $(h_1, h_2, \cdots , h_t)$ of elements from $G$, the cube $C(h_1, h_2, \cdots , h_t)$ is the (multi)set of $2^t$ elements $\{h_1^{\epsilon_1}, h_2^{\epsilon_2}, \cdots , h_t^{\epsilon_t}\}$, with $\epsilon_i \in \{0,1 \}$. Another important feature of the proof is that it works in a very general setting of ``black-box'' groups---it never needs an explicit description of the host group, only the ability to multiply elements and take their inverses. This is a very important paradigm for arguing about groups, and will be used again below.

How does one prove that an element $g$ is {\em not} generated by $S$? It is possible that there is no short ``classical'' proof! This question motivated Babai to define Arthur-Merlin games---a new notion of probabilistic, interactive proofs (simultaneously with Goldwasser, Micali, and Rackoff~\cite{GMR89}, who proposed a similar notion for cryptographic reasons), and showed how non-membership can be certified in this new framework. The impact of the definition of interactive proofs on the theory of computation has been immense, and is discussed in e.g. in the books~\cite{Gol08,ArBa09,Wig17}.

Returning to the generation problem, let us now consider the problem of {\em random generation}. Here we are given $S$, and would like a randomized procedure which will quickly output an (almost) uniform distribution on the subgroup $H$ of $G$ generated by $S$. 
This problem, besides its natural appeal, is often faced by computational group theorists, being a subroutine in many group-theoretic algorithms. In practice often heuristics are used, like the famous ``product replacement algorithm'' and its variants, which often work well in practice (see e.g. the recent~\cite{BaLe12} and references). We will discuss here provable bounds.

It is clear that sufficiently long random words in the elements of $S$ and its inverses will do the job, but just as with certificates, sufficiently long is often prohibitively long. In a beautiful paper, Babai~\cite{Bab91} describes a certain process generating a random program which computes a nearly-uniform element of $H$, and runs in time $n^5 \approx (\log |G|)^5$ steps. It again uses cubes, and works in the full generality of black-box groups. This paper was followed by even faster algorithms with simpler analysis by Cooperman and by Dixon~\cite{Coo02,Dix08}, and the state-of-art is an algorithm whose number of steps is remarkably the same as the length of proofs of generation above---in other words, randomness roughly achieves the efficiency of non-determinism for this problem. Summarizing:

\begin{citedthm}[\cite{Bab91,Coo02,Dix08}]
For every group $G$, there is a {\em probabilistic} program of length $\poly(n) \approx \poly(\log |G|)$ that, given {\em any} generating set $S$ for $G$, produces with high probability a (nearly) uniformly random element of $G$.
\end{citedthm}

\section{Statistical Physics}

The field of statistical physics is huge, and we focus here mainly on connections of statistical mechanics with the theory of computation.
Numerous mathematical models exist of various physical and chemical systems, designed to understand basic properties of different materials and the dynamics of basic processes. These include such familiar models as Ising, Potts, Monomer-Dimer, Spin-Glass, Percolation, etc. A typical example explaining the connection of such mathematical models to physics and chemistry, and the basic problems studied is the seminal paper of Heilmann and Lieb~\cite{HeLi72}.

Many of the problems studied can be viewed in the following general setting. We have a huge (exponential) space of objects called $\Omega$ (these objects may be viewed as the different configurations of a system). Each object is assigned a nonnegative weight (which may be viewed as the ``energy'' of that state). Scaling these weights gives rise to a probability distribution (often called the Gibbs distribution) on  $\Omega$, and to study its properties (phase transitions, critical temperatures, free energy, etc.) one attempts to generate samples from this distribution. Note that if the description of a state takes $n$ bits, then brute-force listing of all probabilities in question is exponentially prohibitive. Thus efficiency of the sampling  procedure is essential to this study.

As $\Omega$ may be highly unstructured, the most common approach to this sampling problem is known as ``Monte Carlo Markov Chain'' (or ``MCMC'') method. The idea is to build a graph on the objects of $\Omega$, with a pair of objects connected by an edge if they are similar in some sense (e.g. sequences which differ only in a few coordinates). Next, one starts from any object, and performs a biased random walk on this graph for some time, and the object reached is the sample produced. In many settings it is not hard to set up the random walk (often called Glauber dynamics or the Metropolis algorithm) so that the {\em limiting} distribution of the Markov chain is indeed the desired distribution. The main question in this approach is {\em when} to stop the walk and output a sample; {\em when} are we close enough to the limit? In other words, how long does it take the chain to converge to the limit? In most cases, these decisions were taken on intuitive, heuristic grounds, without rigorous analysis of convergence time. The exceptions where rigorous bounds were known were typically structured, e.g.\ where the chain was a Cayley graph of a group (e.g. \cite{Ald81,Dia88}).

This state of affairs has changed considerably since the interaction in the past couple of decades with the theory of computation. Before describing it, let us see where computational problems even arise in this field. The two major sources are {\em optimization} and {\em counting}. That the setting above suits many instances of optimization  problems is easy to see. Think of $\Omega$ as the set of solutions to a given optimization problem (e.g.\ the values of certain parameters designed to satisfy a set of constraints), and the weights representing the quality of a solution (e.g.\ the number of constraints satisfied). So, picking at random from the associated distribution favors high-quality solutions. The counting connection is more subtle. Here $\Omega$ represents a set of combinatorial objects one wants to count  or approximate (e.g.\ the set of perfect matchings in a graph, or satisfying assignments to a set of constraints). It turns out that for very general situations of this type, sampling an object (approximately) at random is tightly connected to counting their number; it often allows a recursive procedure to approximate the size of the set~\cite{JVV86}.  An additional observation is that viewing a finite set as a fine discretization of a continuous object (e.g.\ fine lattice points in a convex set) allows one to compute volumes and more generally integrate functions over such domains.

Around 1990, rigorous techniques were introduced~\cite{Ald90,Bro89,SiJe89,DFK91} to analyze the convergence rates of such general Markov chains arising from different approximation algorithms. They establish {\em conductance} bounds on the Markov chains, mainly via {\em canonical paths} or {\em coupling} arguments (a survey of this early work is \cite{JeSi96}). Collaborative work was soon able to formally justify the physical intuition behind some of the suggested heuristics for many models, and moreover drew physicists to suggest such ingenious chains for optimization problems. The field drew in probabilists and geometers as well, and by now is highly active and diverse. We mention two results to illustrate rigorous convergence bounds for important problems of this type. 

\begin{citedthm}[\cite{JSV04}]
The permanent of any nonnegative $n\times n$ matrix can be approximated, to any multiplicative factor $(1+\epsilon)$, in polynomial time in $n/\epsilon$.
\end{citedthm}

The importance of this approximation algorithm stems from the seminal result of Valiant~\cite{Val79a} about the permanent polynomial (that notorious sibling of the determinant polynomial, that looks identical except that the permanent has no signs; for more see~\cite{ShYe10,Wig17}). Valiant proved that the permanent  is {\em universal}, capturing (via efficient reductions) essentially all natural counting problems, including those arising in the statistical physics models and optimization and counting problems above. So, unlike determinant, computing the permanent {\em exactly} is extremely difficult (harder than $\NP$-complete).

\begin{citedthm}[\cite{DFK91}]
The volume of any convex set in $n$ dimensions can be approximated, to any multiplicative factor $(1+\epsilon)$, in polynomial time in $n/\epsilon$.
\end{citedthm}

The volume, besides its intrinsic interest, captures as well natural counting problems, e.g. the number of linear extensions of a given partially ordered set. The analysis of this algorithm, as well as its many subsequent improvements has used and developed purely structural results of independent interest in differential and convex geometry. It also led to generalizations, like efficiently sampling from any log-concave distribution (see the survey~\cite{Vem05}).

Another consequence of this collaboration was a deeper understanding of the relation between {\em spacial} properties (such as phase transitions, and long-range correlations between distant sites in the Gibbs distribution) and {\em temporal} properties (such as speed  of convergence of the sampling or approximately counting algorithms, like Glauber dynamics). This connection (surveyed e.g. in~\cite{DSVW04}) was established by physicists for spin systems since the 1970s. The breakthrough work of Weitz~\cite{Wei06} on the {\em hard core} model gave an {\em deterministic} algorithm which is efficient up to the phase transition, and this was complemented by a hardness result of Sly~\cite{Sly10} beyond the phase transition. These phase transition of computational complexity, at the same point as the phase transition of the Gibbs distribution are striking, and the generality of this phenomenon is still investigated.

More generally, the close similarity between statistical physics models and optimization problems, especially on random instances, is benefitting both sides. Let us mention a few exciting developments. It has unraveled the fine geometric structure of the space of solutions at the phase transition, pinpointing it e.g. for $k$-SAT  in \cite{AcCoRi11}.  At the same time, physics intuition based on such ideas as renormalization, annealing, and replica symmetry breaking, has led to new algorithms for optimization problems, some of them now rigorously analyzed, e.g. as in~\cite{JeSo93}. Others, like one of the fastest (yet unproven) heuristics for such problems as Boolean Satisfiability (which is $\NP$-complete in general) are based on the physics method of ``survey propagation'' of \cite{MPZ02}. Finally, new algorithmic techniques for similar physics and optimization problems, originate from an unexpected source, the  {\em Lovasz Local Lemma} (LLL). The LLL is a probabilistic proof technique for the existence rare events in a probability space. Its efficient versions, formulating it algorithmically as a {\em directed, non-reversible} Markov chains, starting with the works of Moser~\cite{Mos09,MoTa10}, have led to approximate counting and sampling versions for such events (see e.g.~\cite{GJL16}).  A completely different, {\em deterministic} algorithm of Moitra~\cite{Moi16} for the LLL regime (of rare events) promises many more applications: it works even when the solution space (and hence the natural Markov chain) is not connected!

\section{Analysis and Probability}\label{Anal&Prob}

This section gives a taste of a growing number of families of inequalities---large deviation inequalities, isoperimetric inequalities, etc.---that have been generalized beyond their classical origins due to a variety of motivations in the theory of computing and discrete mathematics. Further, the applications sometimes call for {\em stability} versions of these inequalities, namely an understanding of the structures which make an inequality nearly sharp. Here too these motivations pushed for generalizations of classical results and many new ones. Most of the material below, and much more on the motivations, applications and developments in this exciting area of the analysis of Boolean functions, can be found in the book~\cite{Odo14} by  O'Donnell.

The following story can be told from several angles. One is the {\em noise sensitivity} of functions. We restrict ourselves to the Boolean cube endowed with the uniform probability measure, but many of the questions and results extend to arbitrary product probability spaces.
Let $f:\{-1,1\}^n \rightarrow \R$, which we assume is balanced, namely $E[f]=0$. When the image of $f$ is $\{-1,1\}$, we can think of $f$ as a voting scheme, translating the binary votes of $n$ individuals into a binary outcome. One natural desire from such a voting scheme may be {\em noise stability}---that typically very similar inputs (vote vectors) will yield the same outcome. While natural in this social science setting, such questions also arise in statistical physics settings, where natural functions such as bond percolation turn out to be extremely sensitive to noise~\cite{BKS99}. Let us formally define noise stability.

\begin{definition}
Let $\rho \in [0,1]$ be a correlation parameter. We say two vectors $x,y\in \{-1,1\}^n$ are $\rho$-correlated if they are distributed as follows. The vector $x$ is drawn uniformly at random, and $y$ is obtained from $x$ by flipping each bit $x_i$ independently with probability $(1-\rho )/2$. Note that for every $i$ the correlation $E[x_iy_i]=\rho$. The {\em noise sensitivity} of $f$ at $\rho$, $S_{\rho}(f)$, is simply defined as the correlation of the outputs, $E[f(x)f(y)]$.
\end{definition}

It is not hard to see that the function maximizing noise stability is any {\em dictatorship} function, e.g. $f(x)=x_1$, for which $S_\rho(f)=\rho$. But another natural social scientific concern is the {\em influence} of players in voting schemes~\cite{BeLi85}, which prohibits such solutions (in democratic environments). The influence of a single voter\footnote{This seminal paper~\cite{BeLi85}  also studies the influences of coalitions of players, extremely natural in game theory, which arises in and contributes to other areas of computational complexity (including circuit complexity, learning and pseudorandomness), and raises other analytic questions which we will not discuss here.} is the probability with which it can change the outcome given that all other votes are uniformly random (so, in a dictatorship it is 1 for the dictator and 0 for all others). A fair voting scheme should have no voter with high influence. As we define influence for Real-valued functions, we will use the (conditional) {\em variance} to measure a player's potential effect given all other (random) votes.

\begin{definition}
A function $f:\{-1,1\}^n \rightarrow \R$ has influence $\tau$ if for every $i$, $\Var[x_i | x_{-i}] \leq \tau$ for all $i$ (where $x_{-i}$ denotes the vector $x$ without the $i$th coordinate).
\end{definition}

For example, the majority function has influence $O(1/\sqrt{n}$). The question of how small the influence of a balanced function can be is extremely interesting, and leads to a highly relevant inequality for our story (both in content and techniques). As it turns out, ultimate fairness (influence $1/n$ per player) is impossible---\cite{KKL88} show that every function has a player with nonproportional influence, at least $\Omega (\log n / n)$. At any rate, one can ask which of the functions with {\em small} influence is most stable, and it is natural to guess that majority should be the best\footnote{This noise sensitivity tends, as $n$ grows, to $S_\rho (Majority_n) = \frac{2}{\pi} \arcsin \rho$.}.

The conjecture that this is the case, called the {\em Majority is Stablest} conjecture, arose from a completely different and surprising angle---the field of optimization, specifically ``hardness of approximation''.  A remarkable paper~\cite{KKMO07} has shown that this conjecture implies\footnote{Assuming another, complexity-theoretic, conjecture called the ``Unique Games'' conjecture of~\cite{Kho02} (discussed already in the metric geometry section above; see also~\cite{Kho10,Wig17})}
 the optimality of a certain natural algorithm for approximating the {\em maximum cut} of a graph (i.e. the partition of vertices that maximizes the number of edges between them)\footnote{Maximum Cut is a basic optimization problem whose exact complexity is  $\NP$-complete.}. This connection is highly non-trivial, but by now we have many examples showing how the analysis of certain (semidefinite programming-based) approximation algorithms for a variety of optimization problems raise many new isoperimetric questions\footnote{Many over continuous domains, like the unit cube or Gaussian space (see~cite{MoNe12} for one of many examples), where the connection between noise stability and isoperimetry may be ever clearer.}, greatly enriching this field.

The Majority is Stablest conjecture was proved in a strong form by~\cite{MOO10} shortly after it was posed. Here is a formal statement (which actually works for bounded functions).
\begin{citedthm}[\cite{MOO10}]
For every (positive correlation parameter) $\rho \geq0$ and $\epsilon >0$ there exists (an influence bound) $\tau = \tau(\rho, \epsilon)$ such that for every $n$ and every $f:\{-1,1\}^n \rightarrow [-1,1]$ of influence at most $\tau$, $S_\rho (f) \leq S_\rho (Majority_n) + \epsilon$.
\end{citedthm}
The proof reveals another angle on the story---large deviation inequalities and invariance principles. To see the connection, recall the Berry-Esseen theorem~\cite{Fel71}, generalizing the standard central limit theorem to {\em weighted} sums of independent random signs. In this theorem, influences arise very naturally. Consider $\sum_{i=1}^n c_i x_i$. If we normalize the weights $c_i$ to satisfy $\sum_i c_i^2 =1$, then $c_i$ is the influence of the $i$th voter, and $\tau = \max_i |c_i|$. The quality of this central limit theorem deteriorates linearly with the influence $\tau$.  Lindeberg's proof of Berry-Esseen uses an invariance principle, showing that for linear functions, the cumulative probability distribution $Pr[\sum_{i=1}^n c_i x_i \leq t]$ (for every $t$) is unchanged (up to $\tau$), {\em regardless} of the distribution of the variables $x_i$, as long as they are independent and have expectation 0 and variance 1. Thus, in particular, they can be taken to be standard Gaussian, which trivializes the problem, as the weighted sum is a Gaussian as well!
    
To prove their theorem, \cite{MOO10} first observed that also in the noise stability problem, the Gaussian case is simple. If the $x_i, y_i$ are standard Gaussians with correlation $\rho$, the stability problem reduces to a classical result of Borell~\cite{Bor85}: 
that noise stability is maximized by any hyperplane through the origin. Note that here the rotational symmetry of multidimensional Gaussians, which also aids the proof, does not distinguish ``dictator'' functions from majority---both are such hyperplanes. Given this theorem, an invariance principle whose quality depends on $\tau$ would do the job. They next show that it is sufficient to prove the principle only for {\em low degree} multilinear polynomials (as the effect of noise decays with the degree). Finally, they prove this non-linear extension of Berry-Esseen for such polynomials, a form of which we state below. They also use their invariance principle to prove other conjectures, and since the publication of their paper, quite a number of further generalizations and applications were found. 

\begin{citedthm}[\cite{MOO10}]
Let $x_i$ be any $n$ independent random variables with mean 0, variance 1 and bounded 3rd moments. Let $g_i$ be $n$ independent standard Gaussians. Let $Q$ be any degree $d$ multilinear $n$-variate polynomial of influence $\tau$. Then for any $t$,
$$| Pr[Q(x) \leq t] - Pr[Q(g) \leq t] | \leq O(d\tau^{1/d}).$$
\end{citedthm}

We now only seem to be switching gears...
To conclude this section, let me give one more, very different demonstration of the surprising questions (and answers) regarding noise stability and isoperimetry, arising from the very same computational considerations of optimization of hardness of approximation. Here is the question: {\em What is the smallest surface area of a (volume 1) body which tiles $\R^d$ periodically along the integer lattice $\Z^d$?} Namely, we seek a $d$-dimensional volume 1 subset $B\subseteq \R^d$ such that $B + \Z^d = \R^d$, such that its boundary has minimal $(d-1)$-dimensional volume\footnote{Note that the volume of $B$ ensures that the interiors of $B+v$ and $B+u$ are disjoint for any two distinct integer vectors $u,v\in \Z^d$, so this gives a tiling.}. Let us denote this infimum by $s(d)$. The curious reader can stop here a bit and test your intuition, what do you expect the answer to be, asymptotically in $d$? 

Such questions originate from the late 19th century study by Thomson (later Lord Kelvin) of {\em foams} in 3 dimensions~\cite{Tho87a},  further studied, generalized and applied in mathematics, physics, chemistry, material science and even architecture. However, for this very basic question, where periodicity is defined by the simplest integer lattice, it seems that, for large $d$, the trivial upper and lower bounds on $s(d)$ were not improved on for over a century. 
The trivial upper bound on $s(d)$ is provided by the unit cube, which has surface area $2d$. The trivial lower bound on $s(d)$ comes from ignoring the tiling restriction, and considering only the volume - here the unit volume ball has the smallest  surface area, $~\sqrt{2\pi e d}$. Where in this quadratic range does $s(d)$ lie? In particular, can there be {\em ``spherical cubes''}, with $s(d) = O(\sqrt{d})$?

The last question became a central issue for complexity theorists when~\cite{FKO07} related it directly to the important Unique Games conjecture, and optimal inapproximability proofs of combinatorial problems (in particular the maximum cut problem)  discussed above. The nontrivial connection, which the paper elaborates and motivates, goes through attempts to find the tightest version of Raz'~\cite{Raz98a} celebrated parallel repetition theorem\footnote{A fundamental information theoretic inequality of central importance to ``amplification'' of Probabilistically Checkable Proofs (PCPs).}. A limit on how ``strong'' a parallel repetition theorem can get was again provided by Raz~\cite{Raz11}. Extending his techniques~\cite{KORW08} to the geometric setting, resolved the question above, proving that ``spherical cubes'' do exist!

\begin{citedthm}[\cite{KORW08}]
For all $d$, $s(d) \leq \sqrt{4\pi d}$
\end{citedthm}

A simple proof, and various extensions of this result were given subsequently in~\cite{AlKl09}. We note that all known proofs are probabilistic. Giving an explicit construction that might better illustrate how a ``spherical cube'' (even with much worse but non-trivial surface are) looks like, seems like a challenging problem.


\section{Lattice Theory}\label{lattice.sec}

Lattices in Euclidean space are among the most ``universal'' objects in mathematics, in that besides being natural (e.g.\ arising in crystalline structures) and worthy of study in their own right, they capture a variety of problems in different fields such as number theory, analysis, approximation theory, Lie algebras, convex geometry, and more. Many of the basic results in lattice theory, as we shall see, are {\em existential} (namely supply no efficient means for obtaining the objects whose existence is proved), which in some cases has limited progress on these applications.

This section tells the story of one algorithm, of Lenstra, Lenstra, and Lov{\'a}sz~\cite{LLL82}, often called the LLL algorithm, and some of its implications on these classical applications as well as modern ones in cryptography, optimization, number theory, symbolic algebra and more. But we had better define a lattice\footnote{We only define full-rank lattices here, which suffice for this exposition.} first.

Let $B = \{b_1, b_2, \ldots, b_n\}$ be a basis of $\R^n$. Then the {\em lattice} $L(B)$ denotes the set (indeed, Abelian group) of all {\em integer} linear combinations of these vectors, i.e.\ $L(B) = \{ \sum_i z_i b_i \,:\, z_i \in \Z \}$. $B$ is also called a basis of the lattice. Naturally, a given lattice can have many different bases, e.g. the standard integer lattice in the plane, generated by $\{(0,1),(1,0)\}$, is equally well generated by $\{(999,1),(1000,1)\}$. A basic invariant associated with a lattice $L$ is its determinant $d(L)$, which is the absolute value of $\det(B)$ for any basis $B$ of $L$ (this is also the volume of the fundamental parallelpiped of the lattice). For simplicity and without loss of generality, we will assume that $B$ is normalized so that we only consider lattices $L$ of $d(L)=1$.
 
The most basic result about lattices, namely that they must contain {\em short} vectors (in any norm) was proved by Minkowski (who initiated Lattice Theory, and with it, the Geometry of Numbers)~\cite{Min3}.

\begin{citedthm}[\cite{Min3}]
Consider an arbitrary convex set $K$ in $\R^n$ which is centrally symmetric\footnote{Namely, $x\in K$ implies that also $-x \in K$. Such sets are precisely balls of arbitrary norms.} and has volume $> 2^n$. Then, every lattice $L$ (of determinant 1) has a nonzero point in $K$.
\end{citedthm} 

This innocent theorem, which has a simple, but {\em existential} (pigeonhole) proof, turns out to have numerous fundamental applications in geometry, algebra and number theory. Among famous examples  this theorem yields with appropriate choice of norms and lattices, results like Dirichlet's Diophantine approximation theorem and Lagrange's four-squares theorem, and (with much more work) the finiteness of class numbers of number fields (see e.g.\ \cite{PoZa89}).

From now on we will focus on short vectors in the (most natural) Euclidean norm.
A direct corollary of Minkowski's theorem when applying it to the cube $K=[-1,1]^n$ yields: 
 
\begin{corollary}
Every lattice $L$ of determinant 1 has a nonzero point of Euclidean norm at most $\sqrt{n}$.
\end{corollary} 

Digressing a bit, we note that very recently, a century after Minkowski, a strong converse of the above corollary\footnote{Which has to be precisely formulated.} conjectured by Dadush (see~\cite{DaRe16}) for {\em computational} motivation, has been proved in~\cite{ReSt16}. This converse has many structural consequences, on the covering radius of lattices, arithmetic combinatorics, Brownian motion and others. We will not elaborate here on this new interaction of computational complexity and optimization with lattice theory and convex geometry. The papers above beautifully motivate these connections and applications, and the history of ideas and technical work needed for this complex proof.

Returning to Minkowski's corollary for the Euclidean norm, the proof is still existential, and the obvious algorithm for finding such a short vector requires exponential time in $n$. The breakthrough paper~\cite{LLL82} describe the LLL algorithm, an efficient, polynomial-time algorithm, which approximates the length of the shortest vector in any $n$-dimensional lattice by a $2^n$ factor.

\begin{citedthm}[\cite{LLL82}]
There is a polynomial time algorithm, which given any lattice $L$ 
produces a vector in $L$ of Euclidean length at most $2^n$ factor longer than the shortest vector in $L$.
\end{citedthm} 

This exponential bound may seem excessive at first, but the number and diversity of applications is staggering. First, in many problems, the dimension $n$ is a small constant (so the actual input length arises from the bit-size of the given basis). This leads, for instance, to Lenstra's algorithm for (exactly solving) Integer Programming~\cite{Len83} 
in constant dimensions. It also leads to Odlyzko and Riele's refutation~\cite{OdRi85} of Mertens' conjecture about cancellations in the M\"obius function, and to the long list of number theoretic examples in \cite{Sim10}. But it turns out that even when $n$ is arbitrarily large, many problems can be solved in $\poly(n)$-time as well. Here is a list of examples of old and new problems representing this variety, some going back to the original paper~\cite{LLL82}. In all, it suffices that real number inputs are approximated to poly($n$) digits in dimension $n$.
\begin{itemize}
\item {\bf Diophantine approximation}. While the best possible approximation of one real number by rationals with bounded denominator is readily solved by its (efficiently computable) continued fraction expansion, no such procedure is known for {\em simultaneous} approximation. Formally, given a {\em set} of real numbers, say $\{r_1, r_2, \ldots, r_n\}$,  a bound $Q$ and $\epsilon >0$, find integers $q\leq Q$ and $p_1, \ldots, p_n$ such that all $|r_i - p_i/q|\leq \epsilon$. Existentially (using Minkowski), the Dirichlet ``box-principle'' shows that $\epsilon < Q^{1/n}$ is possible. Using LLL, one efficiently obtains $\epsilon < 2^{n^2} Q^{1/n}$ which is meaningful for $Q$ described by $\poly(n)$ many bits.
\item {\bf Minimal polynomials of algebraic numbers}. Here we are given a single real number $r$ and a degree bound $n$, and are asked if there is a polynomial $g(x)$ with integer coefficients, of degree at most $n$ of which $r$ is a root (and also to produce such a polynomial $g$ if it exists). Indeed, this is a special case of the problem above with $r_i = r^i$. While the algorithm only outputs $g$ for which $g(r) \approx 0$, it is often easy to check that it actually vanishes. Note that by varying $n$ we can find the minimal such polynomial.
\item {\bf Polynomial factorization over Rationals}. Here the input is an integer polynomial $h$ of degree $n$, and we want to factor it over $\Q$. The high level idea is to first find an (approximate) root $r$ of $h$ (e.g.\ using Newton's method), feed it to the problem above, which will return a minimal $g$ having $r$ as a root, and thus divides $h$. We stress that this algorithm produces the exact factorization, not an approximate one!
\item {\bf Small integer relations between reals}. Given reals $r_1, r_2, \ldots r_n$, and a bound $Q$, determine if there exist integers $|z_i| <Q$ such that $\sum_i z_i r_i =0$ (and if so, find these integers). As a famous example, LLL can find an integer relation among 
$\arctan(1) \approx 0.785398, \arctan(1/5) \approx 0.197395$ and $\arctan(1/239)
\approx 0.004184$, yielding Machin's formula
$$\arctan(1) - 4 \arctan(1/5) + \arctan(1/239) = 0$$
\item {\bf Cryptanalysis}. Note that a very special case of the problem above (in which the coefficients $z_i$ must be Boolean) is the ``Knapsack problem,'' a  famous $\NP$-complete problem. The point here is that in the early days of cryptography, some systems were based on the assumed ``average case'' hardness of Knapsack. Many such systems were broken by using LLL, e.g.~\cite{Lag84}. LLL was also used to break some versions of the RSA cryptosystem (with ``small public exponents'').
\end{itemize}

It is perhaps a fitting epilogue to the last item that lattices cannot only destroy cryptosystems, but also create them. The problem of efficiently approximating short vectors up to polynomial (as opposed to exponential, as LLL produces) factors is believed to be computationally hard. Here are some major consequences of this assumption. First, Ajtai showed in a remarkable paper~\cite{Ajt96} that such hardness is preserved ``on average'', over a cleverly-chosen distribution of random lattices. This led to a new public-key encryption scheme by Ajtai and Dwork~\cite{AjDw97} based on this hardness, which is arguably the only one known that can potentially sustain quantum attacks (Shor's efficient quantum algorithms can factor integers and compute discrete logarithms~\cite{Sho94}). In another breakthrough work of Gentry~\cite{Gen09}, this hardness assumption is used to devise {\em fully homomorphic} encryption, a scheme which allows not only to encrypt data, but to perform arbitrary computations directly with encrypted data. See more in this excellent survey~\cite{Pei16}.

\section{Invariant Theory}\label{Invariant.sec}

Invariant theory, born in an 1845 paper of Cayley~\cite{Cay45}, is major branch of algebra, with important natural connections to algebraic geometry and representation theory, but also to many other areas of mathematics. We will see some here, as well as some new connections with computational complexity, leading to new questions and results in this field. We note that computational efficiency was always important in invariant theory, which is rife with ingenious algorithms (starting with Cayley's {\em Omega process}), as is evident from the books~\cite{CLO92,Stu08,DeKe15}.

Invariants are familiar enough, from examples like the following. 
\begin{itemize}
\item In high school physics we learn that energy and momentum are preserved (namely, are {\em invariants}) in the dynamics of general physical systems. 
\item In chemical reactions the number of atoms of each element is preserved  as one mixture of molecules is transformed to yield another (e.g. as combining 
Sodium Hydroxide ($NaOH$) and Hydrochloric Acid ($HCl$) yields the common salt Sodium Chloride ($NaCl$) and Water ($H_2O$)). 
\item In geometry, a classical puzzle asks when can a plane polygon  be ``cut and pasted'' along straight lines to another polygon. Here the obvious invariant, {\em area}, is the only one!\footnote{And so, every two polygons of the same area can be cut to produce an {\em identical} (multi)sets of triangles.}. However in generalizing this puzzle to 3-dimensional polyhedra, it turns out that besides the obvious invariant, {\em volume}, there is another invariant, discovered by Dehn\footnote{So there are pairs of 3-dimensional polyhedra of the same volume, which cannot be cut to identical (multi)sets of tetrahedra.}. 
\end{itemize}
More generally, questions about the equivalence of two surfaces (e.g. knots) under homeomorphism, whether two groups are isomorphic, or whether two points are in the same orbit of a dynamical system, etc., all give rise to similar questions and treatment. A canonical way to give negative answers to such questions is through {\em invariants}, namely quantities preserved under some action on an underlying space.

We will focus on invariants of {\em linear groups} acting on {\em vector spaces}. Let us present some notation.
Fix a field $\F$ (while problems are interesting in every field, results mostly work for infinite fields  only, and sometimes just for characteristic zero or algebraically closed ones). 
Let $G$ be a group, and $V$ a representation of $G$, namely an $\F$-vector space on which $G$ acts: for every $g,h\in G$ and $v\in V$ we have $gv \in V$ and $g(hv) = (gh)v$. 

The {\em orbit} under $G$ of a vector (or point) $v\in V$, denoted $Gv$ is the set of all other points hat $v$ can be moved to by this action, namely $\{gv : g\in G\}$. 
Understanding the orbits of a group objects is a  central task of this field. A basic question capturing many of the examples above is, given two points $u,v\in V$, do they lie in the same $G$-orbit, namely if $u\in Gv$. A related basic question, which is even more natural in algebraic geometry (when the field $\F$ is algebraically closed of characteristic zero) is whether the {\em closures}\footnote{One can take closure in either the Euclidean or the Zariski  topology (the equivalence in this setting proved by Mumford~\cite{Mum95}).} of the two orbits intersect, namely if some point in $Gu$ can be approximated arbitrarily well by points in $Gv$. We will return to specific incarnations of these questions.

When $G$ acts on $V$, it also acts on $\F[V]$, the polynomial functions on $V$, also called the {\em coordinate ring} of $V$. In our setting $V$ will have finite dimension (say $m$), and so $\F[V]$ is simply $\F[x_1, x_2, \ldots, x_m] = \F[X]$, the polynomial ring over $\F$ in $m$ variables. We will denote by $gp$ the action of a group element $g\in G$ on a polynomial $p\in \F[V]$.

A polynomial $p(X) \in \F[X]$ is {\em invariant} if it is unchanged by this action, namely for every $g\in G$ we have $gp = p$. All invariant polynomials clearly form a subring of $\F[X]$, denoted $\F[X]^G$, called the {\em ring of invariants} of this action. Understanding the invariants of group actions is the main subject of Invariant Theory. A fundamental result of Hilbert~\cite{Hil2} shows that in our linear setting\footnote{The full generality under which this result holds is actions of  {\em reductive} groups, which we will not define here, but includes all examples we discuss.}, {\em all} invariant rings will be {\em finitely generated} as an algebra\footnote{This means that there is a finite set of polynomials $\{ q_1, q_2, \ldots, q_t \}$ in $\F[X]^G$ so that for every polynomial $p\in \F[X]^G$ there is a $t$-variate polynomial $r$ over $\F$ so that $p = r(q_1, q_2, \ldots, q_t)$.}. Finding the ``simplest'' such generating set of invariants is our main concern here.

Two familiar examples of perfect solutions to this problem follow. 
\begin{itemize}
\item In the first, $G=S_m$, the symmetric group on $m$ letters, is acting on the set of $m$ formal variables $X$ (and hence the vector space they generate) by simply permuting them. Then a set of generating invariants is simply the first $m$ {\em elementary} symmetric polynomials in $X$. 
\item
In the second, $G=SL_n(\F)$, the simple linear group of matrices with determinant 1, is  acting on the vector space $M_n(\F)$ of $n\times n$ matrices (so $m=n^2$), simply by left matrix multiplication. In this case all polynomial invariants are generated by a single polynomial, the determinant of this $m$-variable matrix $X$. 
\end{itemize}

In these two cases, which really supply a complete understanding of the invariant ring $\F[X]^G$, the generating sets are {\em good} in several senses. There are {\em few} generating invariants, they all have {\em low} degree, and they are {\em easy} to compute\footnote{E.g. have {\em small} arithmetic circuits or formulae.}---all these quantities are bounded by a polynomial in $m$, the dimension of the vector space\footnote{There are additional desirable structural qualities of generating sets that we will not discuss, e.g. completely understanding algebraic relations between these polynomials (called {\em syzygies}).} .
In such good cases, one has efficient algorithms for the basic problems regarding orbits of group actions. For example, a fundamental duality theorem of Geometric Invariant Theory~\cite{MFK82} (see Theorem A.1.1), show how generating sets of the invariant ring can be used for the orbit closure intersection problem.

\begin{citedthm}[\cite{MFK82}]
For an algebraically closed field $\F$ of characteristic 0, the following are equivalent for any two $u,v\in V$ and generating set $P$ of the invariant ring $\F[X]^G$.
\begin{itemize}
\item The orbit closures of $u$ and $v$ intersect.
\item For every polynomial $p\in P$, $p(v)=p(u)$.
\end{itemize}
\end{citedthm}

\subsection{Geometric Complexity Theory (GCT)}

We now briefly explain one direction from which computational complexity became interested in these algebraic problems, in work that has generated many new questions and collaboration between the fields. First, some quick background on the main problem of arithmetic complexity theory (see Chapter~\ref{arithmetic.sec} for definitions and more discussion).  In~\cite{Val79a} Valiant defined  arithmetic analogs $\VP$ and $\VNP$ of the complexity classes $\cP$ and $\NP$ respectively, and conjectured that these two arithmetic classes are different (see Conjecture~\ref{vpvnp}). He further proved (via surprising completeness results) that to separate these classes it is sufficient to prove that the {\em permanent} polynomial on $n\times n$ matrices does not project to the {\em determinant} polynomial on $m\times m$ matrices for any $m= \poly(n)$. Note that this is a pure and concrete algebraic formulation of a central computational conjecture.

In a series of papers, Mulmuley and Sohoni introduced {\em Geometric Complexity Theory} (GCT)  to tackle this major open problem\footnote{Origins of using invariant theory to argue computational difficulty via similar techniques go back to Strassen~\cite{Str87}.}. This program is surveyed by Mulmuley here~\cite{Mul12a, Mul11}, as well as in Landsberg's book~\cite{Lan17}. Very concisely, the GCT program starts off as follows. First, a simple ``padding'' of the $n\times n$ permanent polynomial makes it have degree $m$ and act on the entries of an $m\times m$ matrix. Consider the linear group $SL_{m^2}$ action on all entries of such $m\times m$ matrices. This action extends to polynomials in those variables, and so in particular the two we care about: determinant and modified permanent. {\em The main connection is that the permanent projects to the determinant (in Valiant's sense) if and only if the orbit closures of these two polynomials intersect}. Establishing that they do not intersect (for $m = \poly(n)$) naturally leads to questions about finding representation theoretic obstructions to such intersection (and hence, to the required computational lower bound). This is where things get very complicated, and describing them is beyond the scope of this survey. We note that
to date, the tools of algebraic geometry and representation theory were not sufficient even to improve the quadratic bound on $m$ of Theorem~\ref{quadratic-per-det}. Indeed, some recent developments show severe limitations to the original GCT approach (and perhaps guiding it in more fruitful directions); see~\cite{BIP16} and its historical account. Nevertheless, this line of attack (among others in computational complexity) has lead to many new questions in computational commutative algebra and to growing collaborations between algebraists and complexity theorists -- we will describe some of these now.

To do so, we will focus on two natural actions of linear groups on {\em tuples} of matrices, simultaneous conjugation and the left-right action. Both are special cases of {\em quiver representations} (see~\cite{Gab72,DeWe06})\footnote{We will not elaborate on the theory of quivers representation here, but only remark that reductions and completeness occur in this study as well! The left-Right quiver is {\em complete} in a  well defined sense (see \cite{DeMa15}, Section 5). Informally, this means understanding its (semi)-invariants  implies the same understanding of the (semi)-invariants of {\em all} acyclic quivers.}. For both group actions we will discuss the classical questions and results on the rings of invariants, and recent advances motivated by computational considerations.

\subsection{Simultaneous Conjugation}

Consider the following action of $SL_n(\F)$ on $d$-tuples of $n\times n $ matrices. We have $m=dn^2$ variables arranged as $d$ $n\times n$ matrices $X= (X_1, X_2, \ldots, X_d)$. The action of a matrix $Z\in SL_n(\F)$ on this tuple is by simultaneous conjugation, by transforming it to the tuple $(Z^{-1}X_1Z, Z^{-1}X_2Z, \cdots, Z^{-1}X_dZ)$. Now, the general question above, for this action, is which polynomials in the variables $X$ are invariant under this action? 

The work of Procesi, Formanek, Razmyslov, and Donkin~\cite{Pro76,For84,Raz74,Don92} provides a good set (in most aspects discussed above) of generating invariants (over algebraically closed fields of characteristic zero).
The generators are simply the traces of products of length at most $n^2$ of the given matrices\footnote{Convince yourself that such polynomials are indeed invariant.}. Namely the set $$\{ Tr(X_{i_1}X_{i_2}\cdots X_{i_t}) \,:\, t\leq n^2,\, i_j \in [d] \}.$$ These polynomials are explicit, have small degree and are easily computable. The one shortcoming is the {\em exponential} size of this generating set. For example, using it to decide the intersection of orbit closures will only lead to an exponential time algorithm.

By Hilbert's existential Noether's normalization lemma~\cite{Hil2}\footnote{We remark that this is the same foundational paper which proved the {\em finite basis} and {\em Nullstellensatz} theorems. It is interesting that Hilbert's initial motivation to formulate and prove these cornerstones of commutative algebra was the search for invariants of linear actions.} we know that the size of this set of generating invariants can, in principle, be reduced to $dn^2+1$. Indeed, when the group action is on a vector space of dimension $m$, taking $m+1$ ``random'' linear combinations of any finite generating set will result (with probability 1) in a small generating set. However, as we start with an exponential number of generators above, this procedure is both inefficient and also not explicit (it is not clear how to make it deterministic).
One can get an explicit generating set of minimal size deterministically using the Gr\"obner basis algorithm (see~\cite{MaRi11} for the best known complexity bounds) but this will take doubly exponential time in $n$.  

The works above~\cite{Mul12,FoSh13} reduce this complexity to polynomial time! This happened in two stages. First Mulmuley~\cite{Mul12} gave a probabilistic polynomial time algorithm, by cleverly using the structure of the exponentially many invariants above (using which one can obtain sufficiently random linear combinations using only polynomially many random bits and in polynomial time). He then argues that using conditional derandomization results, of the nature discussed in Section~\ref{weakness}, one can derive a deterministic polynomial time algorithm under natural  computational hardness assumptions.  Shortly afterwards, Forbes and Shpilka~\cite{FoSh13} showed that how de-randomized a variant of Mulmuley's algorithm {\em without} any unproven assumption, yielding an unconditional deterministic polynomial time algorithm for the problem! Their algorithm uses the derandomization methodology: very roughly speaking, they first notice that Mulmuley's probabilistic algorithm can be implemented by a very restricted computational model (a certain read-once branching program), and then use an efficient pseudo-random generator for this computational model. Here is one important algorithmic corollary (which can be extended to other quivers).

\begin{citedthm}[\cite{Mul12,FoSh13}]
There is a deterministic polynomial time algorithm to solve the following problem. Given two tuples of {\em rational} matrices $(A_1, A_2, \dots, A_d), (B_1, B_2, \dots, B_d), $, determine if the closure of their orbits under simultaneous conjugation intersect.
\end{citedthm}

It is interesting to remark that if we only consider the orbits themselves (as opposed to their closure), namely ask if there is $Z\in SL_n(\F)$ such that for all $i\in [d]$ we have $Z^{-1}A_iZ = B_i$, this becomes the {\em module isomorphism} problem over $\F$. For this important problem  there is a deterministic algorithm (of a very different nature than above, using other algebraic tools) that can solve the problem over any field $\F$ using only a polynomial number of arithmetic operations over $\F$~\cite{BrLu08}.

\subsection{Left-Right action}

Consider now the following action of two copies, $SL_n(\F) \times SL_n(\F)$ on $d$-tuples of $n\times n $ matrices. We still have $m=dn^2$ variables arranged as $d$ $n\times n$ matrices $X= (X_1, X_2, \ldots, X_d)$. The action of a pair of matrices $Z,W\in SL_n(\F)$ on this tuple is by left-right action, transforming it to the tuple $(Z^{-1}X_1W, Z^{-1}X_2W, \cdots, Z^{-1}X_dW)$. Again, for this action, is which polynomials in the variables $X$ are invariant under this action? 
Despite the superficial similarity to the to simultaneous conjugation, the invariants here have entirely different structure, and bounding their size required different arguments.

The works of~\cite{DeWe00,DoZu01,ScVa01,ANS10} provides an infinite set of generating invariants.
The generators (again, over algebraically closed fields) are determinants of linear forms of the $d$ matrices, with {\em matrix} coefficients of arbitrary dimension. Namely the set
$$\{ \det(C_1 \otimes X_1 + C_2 \otimes X_2 + \dots + C_d \otimes X_d) \,:\, C_i \in M_k(\F), k \in \N \}.$$ 

These generators, while concisely described, fall short on most goodness aspects above, and we now discuss improvements. First, by Hilbert's finite generation, we know in particular that some finite bound $k$ on the dimension of the matrix coefficients $C_i$ exist. A quadratic upper bound $k\leq n^2$ was obtained by Derksen and Makam~\cite{DeMa15} after a  long sequence of improvements described there. Still, there is an exponential number\footnote{Well, a possible infinite number, but it can be reduced to exponential.} of possible matrix coefficients of this size which can be described explicitly, and allowing randomness one can further reduce this number to a polynomial. Thus we e.g. have the following weaker analog to the theorem above regarding orbit closure intersection for this left-right action.

\begin{theorem}
There is a probabilistic polynomial time algorithm to solve the following problem. Given two tuples of {\em rational} matrices $(A_1, A_2, \dots, A_d), (B_1, B_2, \dots, B_d)$, determine if the closure of their orbits under the left-right action intersect.
\end{theorem}

In the remainder we discuss an important special case of this problem, namely when all $B_i=0$, for which a {\em deterministic} polynomial time algorithm was found. While this problem is in commutative algebra, this algorithm surprisingly has implications in analysis and non-commutative algebra, and beyond to 
computational complexity and quantum information theory. We will mention some of these, but let us start by defining the problem.  

For an action of a linear group $G$ on a vector space $V$, define the {\em nullcone} of the action to be the set of all points $v\in V$ such that the the closure of the orbit $Gv$ contains $0$. 
The points in the nullcone are sometimes called {\em unstable}.  The nullcone  of fundamental importance in invariant theory! Some examples of nullcones for actions we have discussed are the following. For the action of $SL_n(\C)$ on $M_n(\C)$ by left multiplication, it is the set of {\em singular} matrices. For the action of $SL_n(\C)$ on $M_n(\C)$ by conjugation, it is the set of {\em nilpotent} matrices. As you would guess (one direction is trivial), the nullcone is precisely the set of points which vanish under all invariant polynomials. Thus if we have a good generating set one can use them to efficiently test membership in the nullcone. However, we are not in this situation for the left-right action. Despite that a derterministic polynomial time algorithm was obtained in~\cite{GGOW15} over the complex numbers, and then a very different algorithm by~\cite{IQS15} which works for all fields. These two algorithms have different nature and properties, and use in different ways the upper bounds on the dimension of matrix coefficients in the invariants.

\begin{citedthm}[\cite{GGOW15,IQS15}]
There is a deterministic polynomial time algorithm, that on a given a tuple of matrices  $(A_1, A_2, \dots, A_d)$ in $M_n(\F)$ determines if it is in the nullcone of the left-right action.
\end{citedthm}

We conclude with some of the diverse consequences of this algorithm. All the precise definitions of the notions below, as well as the proofs, interconnections and the meandering story leading to these algorithms can be found in~\cite{GGOW15,GGOW16}.

\begin{citedthm}[\cite{GGOW15,GGOW16}]
There are deterministic polynomial time algorithms to solve the following problems.
\begin{itemize}
\item The feasibility problem for Brascamp-Lieb ineuqalities, and more generally, computing the optimal constant for each.
\item The word problem over the (non-commutative) free skew field. 
\item Computing the non-commutative rank of  a symbolic matrix\footnote{A matrix whose entries are linear forms in a set of variables}.
\item Approximating the commutative rank of a symbolic matrix to within a factor of two.\footnote{Computing this rank exactly is the PIT problem discussed at the end of Section~\ref{Valiant-Complete}.}.
\item Testing if a completely positive quantum operator is rank-decreasing.
\end{itemize}
\end{citedthm}

\bibliographystyle{alpha}
\bibliography{../avi} {}

\end{document}